\documentclass{article}

\usepackage{arxiv}

\usepackage{times}
\usepackage{xcolor}
\definecolor{lightblue}{rgb}{0.5,0.8,1.0}
\usepackage{soul}
\usepackage[utf8]{inputenc}
\usepackage{amsmath}
\usepackage{stmaryrd}
\usepackage{amssymb}
\usepackage{algorithm}
\usepackage[noend]{algpseudocode}
\usepackage{tikz}
\usetikzlibrary{arrows,%
                petri,%
                topaths}%
\usepackage{tkz-berge}
\usepackage{listings}
\usepackage{amsfonts}
\usepackage{bbding}

\usepackage[textsize=normalsize,textwidth=3cm]{todonotes}

\usepackage{pifont}
\newcommand{\yes}{\ding{52}}
\newcommand{\no}{\ding{56}}

\allowdisplaybreaks 

\usepackage[utf8]{inputenc} 
\usepackage[T1]{fontenc}    
\usepackage{hyperref}       
\usepackage{url}            
\usepackage{booktabs}       
\usepackage{amsfonts}       
\usepackage{nicefrac}       
\usepackage{microtype}      
\usepackage{lipsum}

\title{On the Enactability of Agent Interaction Protocols: Toward a Unified Approach}

\author{
  Angelo Ferrando\thanks{Work supported by EPSRC as part of the ORCA [EP/R026173] and RAIN [EP/R026084] Robotics and AI Hubs.} \\
  Liverpool University\\
  United Kingdom \\
  \texttt{angelo.ferrando@liverpool.ac.uk} \\
  \And
  Michael Winikoff \\
  University of Otago\\
  New Zealand \\
  \texttt{michael.winikoff@otago.ac.nz} \\
  \And
  Stephen Cranefield \\
  University of Otago\\
  New Zealand \\
  \texttt{stephen.cranefield@otago.ac.nz} \\
  \And
  Frank Dignum \\
  Utrecht University \\
  Netherlands \\
  \texttt{F.P.M.Dignum@uu.nl} \\
  \And
  Viviana Mascardi \\
  University of Genova \\
  Italy \\
  \texttt{viviana.mascardi@unige.it} \\
}

\begin{document}
\maketitle

\begin{abstract}
Interactions between agents are usually designed from a global viewpoint. However, the implementation of a multi-agent interaction is distributed. This difference can introduce issues. For instance, it is possible to specify protocols from a global viewpoint that cannot be implemented as a collection of individual agents. This leads naturally to the question of whether a given (global) protocol is enactable. We consider this question in a powerful setting (trace expression), considering a range of message ordering interpretations (what does it mean to say that an interaction step occurs before another), and a range of possible constraints on the semantics of message delivery, corresponding to different properties of underlying communication middleware.
\end{abstract}


\newcommand{\emptyseq}{\epsilon}
\newcommand{\Types}{\mathcal{T}}
\newcommand{\calP}{\mathcal{P}}

\newcommand{\Actions}{\mathcal{A}}
\newcommand{\trans}[1]{\stackrel{{#1}}{\longrightarrow}}
\newcommand{\transmsg}[1]{\stackrel{{#1}}{\Longrightarrow}}
\newcommand{\transmsgrev}[1]{\stackrel{{#1}}{\Longleftarrow}}

\newcommand{\transdist}[1]{\stackrel{{#1}}{\mapsto}}
\newcommand{\isEmpty}{\mathit{\epsilon}}
\newcommand{\acts}{\mathit{tr}}
\newcommand{\ping}{\mathit{ping}}
\newcommand{\pong}{\mathit{pong}}
\newcommand{\msgone}{\mathit{msg_1}}
\newcommand{\msgtwo}{\mathit{msg_2}}
\newcommand{\msgthree}{\mathit{msg_3}}
\newcommand{\ackone}{\mathit{ack_1}}
\newcommand{\acktwo}{\mathit{ack_2}}
\newcommand{\ackthree}{\mathit{ack_3}}
\newcommand{\AltBitOne}{\mathit{AltBit_1}}
\newcommand{\AltBitTwo}{\mathit{AltBit_2}}
\newcommand{\AltBitThree}{\mathit{AltBit_3}}
\newcommand{\Mone}{\mathit{M_1}}
\newcommand{\Mtwo}{\mathit{M_2}}
\newcommand{\fstOptMone}{\mathit{A\Mone}}
\newcommand{\sndOptMone}{\mathit{B\Mone}}
\newcommand{\fstOptMtwo}{\mathit{A\Mtwo}}
\newcommand{\sndOptMtwo}{\mathit{B\Mtwo}}
\newcommand{\Rule}[4]{\scriptstyle{\textrm{({#1})}}{\displaystyle\frac{#2}{#3}}\ #4}
\newcommand{\satype}{\alpha} 
\newcommand{\sa}{a} 

\newcommand{\expansive}[1]{{exp(#1)}}
\newcommand{\abstracted}[1]{{\widetilde{#1}}}
\newcommand{\passiveEvs}[1]{{passive\_events}}
\newcommand{\activeEvs}[1]{{active\_events}}
\newcommand{\passiveInt}[1]{{passive\_int}({#1})}
\newcommand{\activeInt}[1]{{active\_int}({#1})}
\newcommand{\conformance}[4]{{#1} {\conformant_{\langle {#3}, {#4} \rangle}} {#2}}
\newcommand{\conformancestd}[2]{{#1} {\conformant} {#2}}
\newcommand{\substitute}[4]{({#1}\leq_{S}{#2}, {#3}, {#4})}
\newcommand{\completeApply}[1]{complete({#1})}
\newcommand{\complete}[2]{complete({#1})={#2}}

\newcommand{\WAB}{\mathit{WAB}}
\newcommand{\AB}{\mathit{AB}}
\newcommand{\MA}{\mathit{MA}}
\newcommand{\AM}{\mathit{AM}}
\newcommand{\MM}{\mathit{MM}}
\newcommand{\BS}{\mathit{BS}}
\newcommand{\OB}{\mathit{OB}}
\newcommand{\BC}{\mathit{BC}}
\newcommand{\msg}{\mathit{msg}}
\newcommand{\ackType}{\mathit{ack}}
\newcommand{\pmsgone}[1]{\mathit{msg_1^{#1}}}
\newcommand{\pmsgtwo}[1]{\mathit{msg_2^{#1}}}
\newcommand{\packone}[1]{\mathit{ack_1^{#1}}}
\newcommand{\packtwo}[1]{\mathit{ack_2^{#1}}}
\newcommand{\pmsgthree}[1]{\mathit{msg_3^{#1}}}
\newcommand{\packthree}[1]{\mathit{ack_3^{#1}}}
\newcommand{\offer}{\mathit{offer}}
\newcommand{\buy}{\mathit{buy}}
\newcommand{\close}{\mathit{close}}
\newcommand{\closerel}{\triangleright}
\newcommand{\ExtTypes}{\mathcal{T}^+}
\newcommand{\emb}[1]{\mathit{emb}({#1})}
\newcommand{\invemb}[1]{\mathit{emb}^{-1}({#1})}
\newcommand{\extrans}[1]{\stackrel{{#1}}{\rightarrowtail}}
\newcommand{\optrans}[1]{\stackrel{{#1}}{{\rightarrowtail\hspace*{-1em}\longrightarrow}}}
\newcommand{\runset}{\mathcal{R}}
\newcommand{\proj}{\mathit{proj}}
\newcommand{\annot}{\nu}
\newcommand{\pth}{p}
\newcommand{\asa}{b}
\newcommand{\eventSet}{\mathcal{E}}
\newcommand{\interactionSet}{\mathcal{I}}
\newcommand{\agentSet}{\mathcal{A}}
\newcommand{\roleSet}{\mathcal{R}}
\newcommand{\msgSet}{\mathcal{M}}
\newcommand{\agSet}{\mathcal{A}}
\newcommand{\eventTy}{\vartheta}
\newcommand{\messageTy}{\mathit{int}}
\newcommand{\anyEvTy}{\mathit{any}}
\newcommand{\noneEvTy}{\mathit{none}}
\newcommand{\ag}{{ag}}
\newcommand{\agone}{{ag1}}
\newcommand{\agtwo}{{ag2}}
\newcommand{\ev}{ev}
\newcommand{\aevent}{\mathit{ae}}
\newcommand{\run}{\mathit{run}}
\newcommand{\arun}{\mathit{annot\_run}}
\newcommand{\cover}{C}
\newcommand{\shift}[2]{{#1}\downarrow{#2}}

\newcommand{\alice}{\mathit{alice}}
\newcommand{\bob}{\mathit{bob}}
\newcommand{\carol}{\mathit{carol}}
\newcommand{\dave}{\mathit{dave}}
\newcommand{\aamas}{\mathit{aamas}}
\newcommand{\chair}{\mathit{chair}}
\newcommand{\emma}{\mathit{emma}}
\newcommand{\frank}{\mathit{frank}}

\newcommand{\al}{\mathit{a}}
\newcommand{\bo}{\mathit{b}}
\newcommand{\ca}{\mathit{c}}
\newcommand{\da}{\mathit{da}}

\newcommand{\seller}{\mathit{seller}}
\newcommand{\buyer}{\mathit{buyer}}
\newcommand{\tell}{\mathit{tell}}
\newcommand{\price}{\mathit{price}}
\newcommand{\pasta}{\mathit{pasta}}
\newcommand{\pizza}{\mathit{pizza}}
\newcommand{\itemv}{\mathit{item}}
\newcommand{\isItem}{\mathit{isItem}}
\newcommand{\isSeller}{\mathit{isSeller}}
\newcommand{\isBuyer}{\mathit{isBuyer}}
\newcommand{\amount}{\mathit{amount}}
\newcommand{\safe}{\mathit{safe}}
\newcommand{\isEmptyMeth}{\mathit{isEmpty}}
\newcommand{\sizeMeth}{\mathit{size}}
\newcommand{\popMeth}{\mathit{pop}}
\newcommand{\topMeth}{\mathit{top}}
\newcommand{\pushMeth}{\mathit{push}}
\newcommand{\Stack}{\mathit{Stack}}
\newcommand{\NEStack}{\mathit{NEStack}}
\newcommand{\sem}[1]{\llbracket{#1}\rrbracket}
\newcommand{\semThree}[1]{\llbracket{#1}\rrbracket_3}
\newcommand{\Tops}{\mathit{Tops}}

\newcommand{\role}{\mathit{role}}
\newcommand{\hasRole}{\mathit{hasRole}}
\newcommand{\prefixop}{{:}}
\newcommand{\orop}{{\vee}}
\newcommand{\shuffleop}{{|}}
\newcommand{\catop}{{\cdot}}
\newcommand{\allop}{1}
\newcommand{\emptyop}{0}
\newcommand{\andop}{{\wedge}}
\newcommand{\condop}{{\gg}}
\newcommand{\aorb}{\mathit{a\_or\_b}}
\newcommand{\borc}{\mathit{b\_or\_c}}
\newcommand{\TS}{\mathit{TS}}
\newcommand{\restrict}[2]{{#1}_{\mid{#2}}}
\newcommand{\prefixRel}{\ltimes}
\newcommand{\TE}{\mathit{TE}}

\newtheorem{definition}{Def.}[section]
\newtheorem{clai}{Claim}[section]
\newtheorem{conjectur}{Conjecture}[section]

\newcommand{\Guest}{\mathit{Guest}}
\newcommand{\Access}{\mathit{Access}}
\newcommand{\Admin}{\mathit{Admin}}
\newcommand{\RW}{\mathit{RW}}
\newcommand{\login}{\mathit{login}}
\newcommand{\logout}{\mathit{logout}}
\newcommand{\rread}{\mathit{read}}
\newcommand{\wwrite}{\mathit{write}}

\newcommand{\stack}{\mathit{stack}}
\newcommand{\Any}{\mathit{Any}}
\newcommand{\Unsafe}{\mathit{Unsafe}}
\newcommand{\unsafe}{\mathit{unsafe}}

\newcommand{\msgack}{\mathit{msg\_ack}}
\newcommand{\ack}{\mathit{ack}}
\newcommand{\AltBit}{\mathit{AltBit}}

\newcommand{\AP}{\mathit{AP}}
\newcommand{\true}{\mathit{true}}
\newcommand{\until}[2]{{#1}\,U{#2}}
\newcommand{\Inf}{\mathit{Inf}}
\newcommand{\ET}{\mathcal{ET}}

\newcommand{\atomicInt}[3]{{#1}\transmsg{#2}{#3}}
\newcommand{\sendInt}[2]{\transmsg{#1}_{#2}}
\newcommand{\recvInt}[2]{\transmsgrev{#1}_{#2}}

\newcommand{\sendIntA}[3]{{#1}\transmsg{#2}_{(\footnotesize{#3})}}
\newcommand{\recvIntA}[3]{_{(\footnotesize{#1})}\transmsg{#2}{#3}}

\newcommand{\conformant}{{\leq}}
\newcommand{\subs}[2]{{#1}\leq_S{#2}}

\newcommand{\identityMessages}{{Id}_\msgSet}
\newcommand{\identityAgents}{{Id}_\agentSet}

\newcommand{\intlvl}{\emph{Int-Lvl}}
\newcommand{\msglvl}{\emph{Msg-Lvl}}
\newcommand{\commlvl}{\emph{Comm-Lvl}}
\newcommand{\sbs}{\emph{Send before Send}}
\newcommand{\sbr}{\emph{Send before Receive}}
\newcommand{\rbs}{\emph{Receive before Send}}
\newcommand{\rbr}{\emph{Receive before Receive}}
\newcommand{\sbsabbrv}{\emph{SS}}
\newcommand{\sbrabbrv}{\emph{SR}}
\newcommand{\rbsabbrv}{\emph{RS}}
\newcommand{\rbrabbrv}{\emph{RR}}
\newcommand{\synch}{\emph{Synch}}
\newcommand{\rsc}{\emph{RSC}}
\newcommand{\fifo}[2]{\emph{FIFO$_{{#1}-{#2}}$}}
\newcommand{\causal}{\emph{Causal}}
\newcommand{\async}{\emph{Async}}
\newcommand{\stronghappenbeforeop}{\le_{\langle \moi, \commodel \rangle}}
\newcommand{\stronghappenbefore}[4]{{#1}\le_{\langle{#3},{#4}\rangle}{#2}}
\newcommand{\weakhappenbeforeop}{\prec_{\langle \moi, \commodel \rangle}}\newcommand{\weakhappenbefore}[4]{{#1}\prec_{\langle{#3},{#4}\rangle}{#2}}
\newcommand{\stronglocalchoiceop}{\gtrless_{\langle \moi, \commodel \rangle}}
\newcommand{\stronglocalchoice}[4]{{#1}\gtrless_{\langle{#3},{#4}\rangle}{#2}}
\newcommand{\weaklocalchoiceop}{\asymp_{\langle \moi, \commodel \rangle}}
\newcommand{\weaklocalchoice}[4]{{#1}\asymp_{\langle{#3},{#4}\rangle}{#2}}
\newcommand{\first}{\emph{FirstInt}}
\newcommand{\last}{\emph{LastInt}}
\newcommand{\strongenact}[3]{strongly\_enactable({#1},{#2},{#3})}
\newcommand{\strongenactop}{strongly\_enactable}
\newcommand{\weakenact}[3]{weakly\_enactable({#1},{#2},{#3})}
\newcommand{\weakenactop}{weakly\_enactable}
\newcommand{\strongmonitor}[3]{strongly\_monitorable({#1},{#2},{#3})}
\newcommand{\strongmonitorop}{strongly\_monitorable}
\newcommand{\weakmonitor}[3]{weakly\_monitorable({#1},{#2},{#3})}
\newcommand{\weakmonitorop}{weakly\_monitorable}
\newcommand{\semAsync}[3]{\sem{#1}_{#2}^{#3}}
\newcommand{\moi}{\emph{moi}} 
\newcommand{\commodel}{\emph{CM}}
\newcommand{\send}[2]{{#1}{#2}!}
\newcommand{\recv}[2]{{#1}{#2}?}
\newcommand{\sembefore}[1]{\leq_{#1}}
\newcommand{\lang}[2]{\mathcal{L}_{#1}^{#2}}
\newcommand{\interactions}{\mathcal{I}}
\newcommand{\participants}{\mathcal{P}}
\newcommand{\events}{\mathcal{E}_\interactions}
\newcommand{\commodelset}{\emph{ComModel}}
\newcommand{\moiset}{\emph{MOISet}}
\newcommand{\interactionsin}[1]{\interactions({#1})}
\newcommand{\eventsin}[1]{\mathcal{E}_{#1}}
\newcommand{\before}[3]{{#1}\prec_{#2}{#3}}
\newcommand{\beforems}[4]{{#3}\prec_{#1}^{#2}{#4}}
\newcommand{\beforemsop}{\prec_{\moi}^{u}}
\newcommand{\causalrel}[1]{\prec_{\causal}^{#1}}
\newcommand{\countasdef}{countAs:\moi\times\commodel\trans{}\mathcal{P}(\moi)}
\newcommand{\countasop}{countAs}
\newcommand{\countas}[3]{countAs({#1}, {#2}) = {#3}}
\newcommand{\countasappl}[2]{countAs({#1}, {#2})}
\newcommand{\strconstrmsdef}{satisfy_{s}:\moi\times\interactions\times\interactions\trans{}\mathcal{B}}
\newcommand{\strconstrmsop}{satisfy_{s}}
\newcommand{\strconstrms}[4]{satisfy_{s}({#1},{#2},{#3})={#4}}
\newcommand{\strconstrmsappl}[3]{satisfy_{s}({#1},{#2},{#3})}
\newcommand{\weakconstrmsdef}{satisfy_{w}:\moi\times\interactions\times\interactions\trans{}\mathcal{B}}
\newcommand{\weakconstrmsop}{satisfy_{w}}
\newcommand{\weakconstrms}[4]{satisfy_{w}({#1},{#2},{#3})={#4}}
\newcommand{\weakconstrmsappl}[3]{satisfy_{w}({#1},{#2},{#3})}
\newcommand{\langinteractions}{\lang{\interactions}{}}
\newcommand{\langevents}{\lang{\events}{}}
\newcommand{\langeventsin}[1]{\lang{\mathcal{E}_{#1}}{}}
\newcommand{\interaction}{{i}}
\newcommand{\inttrace}{{I}}
\newcommand{\event}{{e}}
\newcommand{\eventtrace}{{E}}
\newcommand{\msgsemantic}{MOI}

\newcommand{\ags}[1]{ag({#1})}

\section{Introduction}
\label{sec:intro}

In order to organise her staying in Montreal, Alice books an apartment from Bob via the online platform AIPbnb. AIPbnb policy states that owners cannot interact with each other, users can interact with owners only via the platform, and if a user finds a better solution for her accommodation, she must  cancel the previous one {\em before} she makes a new reservation for the same dates, otherwise she will be charged for one night there. 
When Alice discovers that Carol rents a cheaper and larger apartment, she decides to cancel the reservation of Bob's apartment and book Carol's one. 
This situation can be represented by the global Agent Interaction Protocol
$\mathit{modifyRes} = Alice \transmsg{Canc} Bob ~\cdot~  Alice \transmsg{Res} Carol$ where $a1 \transmsg{M} a2$
models the interaction between $a1$ and $a2$ for exchanging message $M$, ``$\cdot$'' models interaction concatenation, and $Canc$ and $Res$ are sent to the recipients by using the AIPbnb platform as required. 
Alice believes that the above protocol correctly meets AIPbnb policy, but she is charged for one night in Bob's apartment by AIPbnb: Carol received Alice's request before Bob received the cancellation, and this violates the policy.
What went wrong is the {\em interpretation} of ``before''. To Alice, it meant that she should send $Canc$ before she sent $Res$, while for AIPbnb it (also) meant that Bob should receive $Canc$ before Carol received $Res$. This ambiguity would have had no impact on Alice if the physical {\em communication model} underlying AIPbnb guaranteed that between the sending and receiving stages of an interaction, nothing could happen. 
However, if the communication model provides weaker or no guarantees,
it may happen that a message sent before another, is delivered after.

This simple example shows that enacting the respect of a global protocol without a clear semantics of the ``before'' meaning, without guarantees from the platform implementation on message delivery order, and without hidden communications between the participants (``covert channels''), may not be possible. Many real situations can be resorted to this one: for example, a citizen must wait for the bank to have received (and processed) the request for adding some money to a new, empty account, before sending a request to move that amount to another account, otherwise he can go in debt.

Global protocols are modelled using many different formalisms including global types \cite{DBLP:conf/forte/CastagnaDP11}, Petri Nets \cite{Peterson:1977:PN:356698.356702}, WS-CDL \cite{wscdl}, 
AUML~\cite{Huget05},  Statecharts~\cite{Harel1987231}, and causal logic \cite{DBLP:journals/ai/GiunchigliaLLMT04}. In each of these formalisms the enactability problem, that we define as ``by executing the localised versions of the protocol implemented by each participant, the global protocol behaviour is obtained, with no additional communication'',
has been addressed in some form. Despite their diversity, however, most of these formalisms do not support protocol concatenation and recursion, which are needed to achieve a high expressivity: their expressive power is limited to regular languages. 

Moreover, although -- from an operational point of view -- these approaches agree on the intuition that a global protocol is enactable if the composition of the local protocols, obtained by projecting the global one onto each participant, behaves exactly in the same way as the global protocol,
the {\em semantic definition} of enactability is far from being standard and sometimes is also more restrictive than necessary: some protocols will be classified as not enactable, while (under suitable conditions) they could be enacted. 

The intended {\em message ordering} and the {\em communication model} of the infrastructure in which the agents will be implemented and run are never taken into consideration together. As shown in the example above these two elements are effectively two sides of the same coin which must be both modeled for providing a precise and generally applicable definition of enactability.

In a similar way, the need to associate the protocol with a {\em decision structure} to enforce consistent choices, is recognised as a necessity and suitably addressed by \cite{DBLP:conf/www/QiuZCY07} only, and not in conjunction with the other issues that affect enactability.  

Finally, the availability of a {\em working prototype} to check the enactability of global protocols under message ordering and communication models is usually disregarded in the literature.  

In this paper we provide a semantic characterisation of enactability which integrates {\em message ordering} and {\em communication model} in a unified framework, along with {\em decision structures}. This combination prevents unnecessary restrictions from the definition, which is as general as possible and suitable for highly expressive protocol representation languages like Trace Expressions \cite{frankDeBoer2015}. We also developed a working prototype in Haskell for enactability checks, which is one key benefit of out approach. 


\section{Background}
\label{sec:backRel}


\paragraph{Trace Expressions.}\label{sec:traceexpressions}
Trace expressions \cite{frankDeBoer2015} are a compact and expressive formalism inspired by global types \cite{DBLP:conf/dalt/AnconaDM12} and then extended and exploited in different application domains \cite{DBLP:conf/atal/AnconaFM17,DBLP:conf/atal/FerrandoAM17,DBLP:conf/atal/FerrandoDA0M18,DBLP:conf/ecoop/AnconaFFM17,DBLP:conf/atal/FerrandoAM16}. 
Trace Expressions, initially devised for runtime verification of multiagent systems, are able to define languages that are more than context free.
 
A trace expression $\tau$ denotes a set of possibly infinite event traces, and is defined on top of the following operators:\footnote{Binary operators associate from left, and are listed in decreasing order of precedence, that is, the first operator
has the highest precedence.}
\begin{itemize}
\item $\emptyseq$ (empty trace), denoting the singleton set $\{\langle \rangle\}$ containing  the empty event trace $\langle \rangle$.
\item $M$ (event), denoting a singleton set $\{\langle M \rangle\}$ containing the event trace $\langle M \rangle$.
\item $\tau_1\catop\tau_2$ (\emph{concatenation}), denoting the set of all traces obtained by concatenating the traces of $\tau_1$ with
those of $\tau_2$. 
\item $\tau_1\andop \tau_2$ (\emph{intersection}), denoting the intersection of the traces of $\tau_1$ and $\tau_2$. 
\item $\tau_1\orop \tau_2$ (\emph{union}), denoting the union of the traces of $\tau_1$ and $\tau_2$. 
\item $\tau_1\shuffleop \tau_2$ (\emph{shuffle}), denoting the union of the sets obtained by shuffling each trace of $\tau_1$ with each trace of
  $\tau_2$ (see \cite{DBLP:journals/iandc/BrodaMMR18} for a more precise definition).
\end{itemize}

Trace expressions are cyclic terms, thus they can support recursion without introducing an explicit construct.

As customary, the operational semantics of trace expressions, defined in \cite{DBLP:conf/birthday/AnconaFM16}, is specified by a transition relation 
$\delta\subseteq\Types\times\eventSet\times\Types$, where $\Types$ and $\eventSet$ denote the set of trace expressions
and of events, respectively. We do not present all the transition rules for space constraints. They are standard ones which state, for example, that  $\delta(\ev \cdot \tau, \ev, \tau)$ (the protocol whose state is modelled by $\ev \cdot \tau$ can move to state $\tau$ if $\ev$ occurs), and that  $\delta(\tau_1 \lor \tau_2, \ev, \tau)$ if $\delta(\tau_1, \ev, \tau)$ (if the protocol whose state is modelled by $\tau_1$ can move to state $\tau$ if $\ev$ occurs, then also the protocol whose state is modelled by $\tau_1  \lor \tau_2$ can).

The denotational semantics is defined as follows: \label{sec:standardsemantics}
\begin{eqnarray*}
\sem{\emptyseq} &=& \{ \langle \rangle \} \\
\sem{M} &=& \{\langle M \rangle \} \\
\sem{\tau_1 \cdot \tau_2} &=& \{ t_1 \circ t_2 | t_1 \in \sem{\tau_1} \land t_2 \in \sem{\tau_2} \} \\
\sem{\tau_1 \land \tau_2} &=& \sem{\tau_1} \cap \sem{\tau_2} \\
\sem{\tau_1 \lor \tau_2} &=&  \sem{\tau_1} \cup \sem{\tau_2} \\
\sem{\tau_1 | \tau_2} &=& \{ z \; |\;   t_1 \in \sem{\tau_1} \land t_2 \in \sem{\tau_2} \land z \in t_1 \bowtie t_2\} \\
\end{eqnarray*}
Where $t_1 \bowtie t_2$ is the set of all interleavings of  $t_1$ and $t_2$, and $\circ$ is concatenation over sequences.

Events can be in principle of any kind. In this paper, we will limit ourselves to consider \emph{interaction} and \emph{message} events.

An interaction has the form $a\transmsg{M}b$ and gives information on the protocol from the global perspective, collapsing sending and receiving. We say that $\tau$ is an interaction protocol if all the events therein are interactions. Interaction protocols take other names in other communities, such as Interaction Oriented Choreography \cite{DBLP:conf/sefm/LaneseGMZ08} in the Service Oriented Community, and global type in the community working on process calculi and types \cite{DBLP:conf/forte/CastagnaDP11}.

Message events have the form $\send{a}{M}$ ($a$ sends $M$) and $\recv{b}{M}$ ($b$ receives $M$). They model actions that one agent can execute, hence taking a local perspective. A trace expression where all events are messages will be named a message protocol throughout the paper. Message protocols have different names in different communities, such as Process Oriented Choreography \cite{DBLP:conf/sefm/LaneseGMZ08} and ``local
type'' or ``session type'' in the global type community \cite{DBLP:conf/esop/HondaVK98,DBLP:conf/parle/TakeuchiHK94}.

\paragraph{Communication Models.}
\label{sec:commMod}

Given that in our proposal we explicitly take the communication model supported by the MAS infrastructure into account, we provide a summary of communication models based on \cite{DBLP:journals/fac/ChevrouHQ16}. We use CM0 to CM6 to identify them in a compact way. 

\noindent {\bf CM0: Synchronous Communication}. Sending and receiving are synchronised: the sender cannot send if the receiver is not ready to receive.

\noindent {\bf CM1: Realisable with Synchronous Communication (RSC)}. After a communication
transition consisting of a send event of a message, the only possible communication transition is the receive event
of this message. This asynchronous model is the closest one to synchronous communication and can be implemented with a 1-slot unique buffer shared by all agents.

\noindent {\bf CM2: FIFO n-n communication}. Messages are globally ordered and are delivered in their emission order: if sending of $M_1$ takes place before sending of $M_2$, then reception of $M_1$ must take place before reception of $M_2$. This model can be implemented by means of a shared centralised
object, such as unique queue.

\noindent {\bf CM3: FIFO 1-n communication}. Messages from the same sender are delivered in the order in which they were sent. It can be implemented by giving each agent a unique queue where it puts its outgoing messages. Destination peers fetch messages from this queue. 

\noindent {\bf CM4: FIFO n-1 communication}. A send event is implicitly and globally ordered with regard to all other sending actions toward the same agent. This means that if 
agent $b$ receives $M_1$ (sent by agent $a$) and later it receives $M_2$ (sent by agent $c$), $b$ knows that the sending of $M_1$
occurred before the sending of $M_2$ in the global execution order, even if there is no causal path between the two sending actions. The implementation of this model can, similarly to FIFO 1-n, be done by providing each agent with a queue: messages are sent by putting them into the queue of the recipient agent.

\noindent {\bf CM5: Causal}. Messages are delivered according to the causality of their emissions \cite{Lamport:1978:TCO:359545.359563}: if a message $M_1$ is
causally sent before a message $M_2$  then
an agent cannot get $M_2$ before $M_1$. An implementation of this model requires the sharing of the causality relation.

\noindent {\bf CM6: Fully Asynchronous}. No order on message delivery is imposed. Messages can overtake others or be arbitrarily delayed. The implementation is usually modelled by a bag.

\paragraph{Message Ordering.}
\label{sec:moiMod}

The statement ``one interaction comes before another'' is ambiguous, as exemplified in Section \ref{sec:intro}. This ambiguity has been recognised by some authors who suggested how to interpret message ordering, when moving from the interaction (global) level to the message (local) level. In this section we summarise and compare the proposals by Lanese, Guidi, Montesi and Zavattaro \cite{DBLP:conf/sefm/LaneseGMZ08} and that by Desai and Singh \cite{DBLP:conf/aaai/DesaiS08}. 

To identify the interpretations, we will use the acronyms used in \cite{DBLP:conf/aaai/DesaiS08} when available, and our own acronyms otherwise. The starting point for interpreting message ordering is the interaction protocol $\tau = a\transmsg{M_1}b \catop c\transmsg{M_2}d$. For the sake of clarity, we denote $\send{a}{M_1}$ with $s1$, $\recv{b}{M_1}$ with $r1$, $\send{c}{M_2}$ with $s2$, and $\recv{d}{M_2}$ with $r2$; we characterise the message ordering interpretations by the traces of messages that respect them.

\noindent {\bf RS}: a message send must be followed immediately by the corresponding receive, so w.r.t. $\tau$, $M_1$ must be received before $M_2$ is sent. The set of traces that respect this model is 
$\{ s1~r1~s2~r2 \}$. This interpretation is named {\em RS (receive before send)} in \cite{DBLP:conf/aaai/DesaiS08} and {\em disjoint semantics} in \cite{DBLP:conf/sefm/LaneseGMZ08}. 

\noindent {\bf SS}: $M_1$ is sent before $M_2$ is, and there are no constraints on the delivery order. The set of traces that respect this model is $\{ s1~r1~s2~r2,$ $s1~s2~r1~r2,$ $s1~s2~r2~r1 \}$. This interpretation is named {\em SS (send before send)} in \cite{DBLP:conf/aaai/DesaiS08} and {\em sender semantics} in \cite{DBLP:conf/sefm/LaneseGMZ08}.

\noindent {\bf RR}: $M_1$ is received before $M_2$ is, and there are no constraints on the sending order. The set of traces that respect this model is $\{ s1~r1~s2~r2, s1~s2~r1~r2, s2~s1~r1~r2 \}$. This interpretation is named {\em RR (receive before receive)} in \cite{DBLP:conf/aaai/DesaiS08} and {\em receiver semantics} in \cite{DBLP:conf/sefm/LaneseGMZ08}.

\noindent {\bf RR \& SS}: this combines the requirements of {\bf RR} and of {\bf SS}: $M_1$ is sent before $M_2$ is sent and also $M_1$ is received before $M_2$ is received. The set of traces that respect this model is $\{ s1~r1~s2~r2,   s1~s2~r1~r2 \}$: both $s1$ comes before $s2$ (``coming before'' according to the senders), and $r1$ comes before $r2$ (``coming before'' according to the receivers). This interpretation is named {\em sender-receiver semantics} in \cite{DBLP:conf/sefm/LaneseGMZ08}. 

\noindent {\bf SR}: $M_1$ is sent before $M_2$ is received. The set of traces that respect this model is $\{ s1~r1~s2~r2,$ $s1~s2~r1~r2,$ $s1~s2~r2~r1,$ $s2~s1~r1~r2,$ $s2~s1~r2~r1 \}$. This interpretation is named {\em SR (send before receive)} in \cite{DBLP:conf/aaai/DesaiS08}.

It is easy to see that the following inclusions among asynchronous models hold: {\bf RS}  $\subset$ {\bf RR \& SS}  $\subset$ {\bf SS} $\subset$ {\bf SR}  and  {\bf RS}  $\subset$ {\bf RR \& SS}  $\subset$ {\bf RR} $\subset$ {\bf SR}. The {\bf SS} and {\bf RR} interpretations are not comparable.   
In the remainder of this paper we consider only the four interpretations defined by Desai \& Singh, i.e.~we do not consider ``RR \& SS''.

\section{Defining Enactability using a Semantic Approach}\label{sec:semantics}

\paragraph{Basic Notation.}\label{sec:sembg}

In the following let  
$\commodelset= \{CM1, CM2, CM3, CM4, CM5,$ $ CM6\}$
be the set of possible (asynchronous) communication models, and $\moiset=$ $\{\sbsabbrv$, $\sbrabbrv$, $\rbsabbrv$, $\rbrabbrv$ $\}$ the set of possible message order interpretations that can be imposed.

We also define $\agentSet=\{a, b, c, d, a_1, a_2, \ldots , a_n\}$ to be the set of agents involved in the interaction protocol. 

Recall that we consider both interaction and message protocols. 
When we say that $\tau$ is an \emph{interaction} protocol, we mean that the protocol represents sequences of \emph{interactions}. The set of traces recognized is obtained following the semantics defined in Section~\ref{sec:traceexpressions}, and for an interaction protocol $\tau$ we have that\footnote{We use ``$\in$'' to also denote  membership of an item in a sequence.} $\inttrace \in \sem{\tau} \implies \forall_{\interaction \in \inttrace}.\interaction\in\interactionsin{\tau}$, where we define
$\interactionsin{\tau}$  to be the set of interactions involved in the interaction protocol $\tau$. We also define $\interactions$ to be the set of all possible interactions events.
Similarly, when $\tau$ is a \emph{message} protocol (rather than an interaction protocol), it represents sequences of send and receive events of the form $\send{a}{M}$ (send event) and $\recv{b}{M}$ (receive event), and given a particular set of possible interactions $\interactions$, we define $\events$ to be the corresponding set of events: $$\events = \{ \send{a}{M} | \exists_{b\in\agentSet}.a\transmsg{M}b\in\interactions \} \cup  \{ \recv{b}{M} | \exists_{a\in\agentSet}.a\transmsg{M}b\in\interactions \} $$  
In a message protocol $\tau$ we have that $\eventtrace \in \sem{\tau} \implies \forall_{\event\in\eventtrace}.\event \in \eventsin{\interactionsin{\tau}}$. 
Given a message protocol $\tau$ we also define $\eventsin{}(\tau)$ to be the set of events that occur in the protocol.

Next, we define the language of traces for interaction protocols and message protocols. For interaction protocols, the set of all possible traces is defined to be: $\langinteractions = \interactions^*\cup\interactions^\omega$. For message protocols the definition is somewhat more complex, since there is a relationship between a send and a receive event. Specifically, the set of all possible traces of events is constrained so that a message being received must be preceded by that message having been sent.
We also constrain the set so that each message can be sent at most once, and received at most once (i.e.~message names are unique). The assumption is made by most authors, see \cite{DBLP:journals/fac/ChevrouHQ16} for example, and it is considered as a harmless one; we can integrate many elements to the notion of ``message name'', such as content, protocol id, conversation id, etc, to discriminate between messages at design time.
Formally: 
\begin{eqnarray*}
\langevents &=& \{ \eventtrace \in \events^*\cup\events^\omega  \; |  \\
& & \hspace*{-12mm} (\forall_{i,j\in dom(\eventtrace)}.\eventtrace[i] = \send{a}{M} \wedge \eventtrace[j] = \send{a}{M} \implies i = j) \wedge \\
& & \hspace*{-12mm}(\forall_{i,j\in dom(\eventtrace)}.\eventtrace[i] = \recv{b}{M} \wedge \eventtrace[j] = \recv{b}{M} \implies i = j) \wedge \\
& & \hspace*{-12mm} (\forall_{i\in dom(\eventtrace)}.\eventtrace[i] = \recv{b}{M} \implies (\exists_{j\in dom(\eventtrace)}.\eventtrace[j] = \send{a}{M} \wedge j < i)) 
\end{eqnarray*}

\paragraph{Message Order Interpretation (MOI).}
An interaction protocol $\tau$ defines orderings between messages $M_i$, whereas a message protocol deals in \emph{events} (sending and receiving). If a protocol says that $M_1$ comes before $M_2$, how should we interpret this in terms of events? Should sending $M_1$ come before sending $M_2$, or does it mean that receiving $M_1$ should occur before receiving $M_2$? The \emph{message ordering interpretation} (MOI) specifies this. As discussed earlier, we follow prior work in considering four (natural) interpretations ($\sbsabbrv$, $\sbrabbrv$, $\rbsabbrv$, and $\rbrabbrv$).
We formalise this by defining a variant semantics that takes an \emph{interaction} protocol $\tau$ and returns its semantics in terms of \emph{events} rather than interactions. The possible sequences of events are constrained: given a situation where $\tau$ specifies that $M_1$ must occur before $M_2$, we constrain the possible sequence of events with the appropriate constraint on events corresponding to the selected MOI.

\begin{definition}[Order on interactions in a trace]
Let $\inttrace\in\langinteractions$ be a trace of interaction events, $\eventtrace\in\langevents$ be a trace of send and receive events, $\moi\in\moiset$ a message ordering interpretation, and $a \transmsg{M_1} b \in \interactions$, $c \transmsg{M_2} d \in \interactions$ two interactions. Abbreviating $a \transmsg{M_1} b$ as $I_1$ and $c \transmsg{M_2} d$ as $I_2$, we define an order on $M_1$ and $M_2$ for $\moi$ in $\eventtrace$ as follows:\\
\noindent$\beforems{\sbsabbrv}{\eventtrace}{I_1}{I_2} \triangleq \before{\send{a}{M_1}}{\eventtrace}{\send{b}{M_2}}$\\
\noindent$\beforems{\sbrabbrv}{\eventtrace}{I_1}{I_2} \triangleq \before{\send{a}{M_1}}{\eventtrace}{\recv{d}{M_2}}$\\
\noindent$\beforems{\rbsabbrv}{\eventtrace}{I_1}{I_2} \triangleq \before{\recv{b}{M_1}}{\eventtrace}{\send{b}{M_2}}$\\
\noindent$\beforems{\rbrabbrv}{\eventtrace}{I_1}{I_2} \triangleq \before{\recv{b}{M_1}}{\eventtrace}{\recv{d}{M_2}}$ \\
where $\before{e_1}{\eventtrace}{e_2}\triangleq \exists_{i,j\in dom(\eventtrace)}.\eventtrace[i] = e_1 \wedge \eventtrace[j] = e_2 \wedge i \leq j$
\end{definition}

Formalising the MOI is not as simple as it might seem. An obvious approach that does not work is to compute the semantics of the interaction protocol $\tau$, and then map each sequence $\inttrace \in \sem{\tau}$ to a set of message event traces. This does not work because  the trace is linear, and therefore a total order, whereas a protocol can specify a partial order. 
An illustrative example is $\tau = (M_1 \cdot M_2) \; | \; M_3$. This simple protocol has three sequences of interactions: $\{ \langle M_1, M_2, M_3 \rangle, \langle M_1, M_3, M_2 \rangle, \langle M_3, M_1, M_2 \rangle \}$. Assume an RS message ordering interpretation, then each of the message sequences corresponds to exactly one sequence of events, giving\footnote{For readability we use $s(M)$ and $r(M)$ to abbreviate sending and receiving message $M$, eliding the identity of the agents involved.}
$\{ 
\langle s(M_1), r(M_1), s(M_2), r(M_2), s(M_3), r(M_3)\rangle,$ $\langle s(M_1), r(M_1), $ $ s(M_3), r(M_3),$ $ s(M_2), r(M_2)  \rangle, \langle s(M_3), r(M_3), s(M_1), r(M_1),  s(M_2), $ $  r(M_2) \rangle  \}$.  However, the protocol does not specify any constraint on $M_3$, so should also allow other interpretations where the occurrences of $s(M_3)$ and $r(M_3)$ are not constrained relative to the other events, for example $ \langle s(M_1), r(M_1), s(M_3), s(M_2), r(M_2),$ $  r(M_3)  \rangle $.

Instead, we define a variant semantics, which is compositional. 
The semantics follow the standard semantics (Section~\ref{sec:standardsemantics}) with a few exceptions. Firstly, the semantics of an interaction $I$ is given as the sequence of sending the message, followed by receiving it (denoted respectively $s(I)$ and $r(I)$). Secondly, the semantics for a sequence $\tau_1 \cdot \tau_2$ is given by taking the semantics of $\tau_1$ and of $\tau_2$. These are then combined by interleaving them (rather than simply concatenating them), but with the constraint that the result must satisfy the appropriate MOI constraint ($\beforems{\sbsabbrv}{\eventtrace}{I_1}{I_2}$) for all possible final messages of $\tau_1$ ($I_1$) and all possible initial messages of $\tau_2$ ($I_2$). Determining initial and final messages is itself somewhat complex, and is done using partially ordered sets.

A partially ordered set (poset) is a pair $(E, <)$ where $E$ is the set of elements (in this case send and receive events) and $<$ is a binary relation on $E$. We define the union operator to act piecewise on posets, and to take the transitive closure of the resulting relation, i.e. $(E_1, <_1) \cup (E_2, <_2) = (E_1 \cup E_2 , (<_1 \cup <_2)^*)$. 
We can then define the $\mathrm{poset}$ of an interaction protocol as follows
\begin{eqnarray*}
\mathrm{poset}(\epsilon) &=& (\varnothing, \varnothing)\\
\mathrm{poset}(I) &=& (\{I\},\varnothing)\\
\mathrm{poset}(\tau_1 \land \tau_2) &=& \mathrm{poset}(\tau_1) \cup \mathrm{poset}(\tau_2) \\ 
\mathrm{poset}(\tau_1 \mathop{|} \tau_2) &=& \mathrm{poset}(\tau_1) \cup \mathrm{poset}(\tau_2) \\
\mathrm{poset}(\tau_1 \lor \tau_2) &=& \mathrm{poset}(\tau_1) \cup \mathrm{poset}(\tau_2) \\
\mathrm{poset}(\tau_1 \cdot \tau_2) &=& \mathrm{poset}(\tau_1) \cdot \mathrm{poset}(\tau_2) \\
(E_1, <_1) \cdot (E_2, <_2) &=& (E_1 \cup E_2 , <_1 \cup <_2 \cup  \{ (x,y) \; | \\ & & x \in \max(E_1,<_1) \land y \in \min(E_2,<_2) \})
\end{eqnarray*}

Where we define a sequence of two posets $(E_1, <_1) \cdot (E_2, <_2)$ by collecting the orderings of each of $E_1$ and $E_2$, and adding additional ordering constraints between the maximal elements of $E_1$ and the minimal elements of $E_2$. We can now proceed to define $\sem{\tau}_{\moi}$.
\begin{eqnarray*}
\semAsync{\emptyseq}{\moi}{} &=& \{ \emptyseq \}\\
\semAsync{I}{\moi}{} &=& \{ \langle s(I), r(I) \rangle \} \\
\semAsync{\tau_1 \land \tau_2}{\moi}{} &=& \semAsync{\tau_1}{\moi}{} \cap  \semAsync{\tau_1}{\moi}{} \\
\semAsync{\tau_1 \cdot  \tau_2}{\moi}{} &=& \{ t \,|\,  t_1 \in \semAsync{\tau_1}{\moi}{}  \land t_2 \in \semAsync{\tau_2}{\moi}{}  \land t \in t_1 \bowtie t_2 \land {} \\ 
	& & \hspace*{3mm} \forall I_1 \in \mathrm{max}(\mathrm{poset}^{}(\tau_1)), \\ 
	& & \hspace*{6mm} \forall I_2 \in \mathrm{min}(\mathrm{poset}^{}(\tau_2)):  \beforems{\moi}{t}{I_1}{I_2} \} \\ 
\semAsync{\tau_1 \lor \tau_2}{\moi}{} &=& \semAsync{\tau_1}{\moi}{} \cup  \semAsync{\tau_1}{\moi}{} \\
\semAsync{\tau_1 | \tau_2}{\moi}{} &=& \{ z \; | \;   t_1 \in \semAsync{\tau_1}{\moi}{} \land t_2 \in \semAsync{\tau_2}{\moi}{} \land z \in t_1 \bowtie t_2\} 
\end{eqnarray*}

Where $t_1 \bowtie t_2$ is the set of all interleavings of  $t_1$ and $t_2$.

\paragraph{Communication Model Semantics.}\label{sec:communicationmodelsemantics}

We formalise the defined communication model semantics by defining for each communication model $CMi$ a corresponding language of event traces that incorporates the appropriate restriction, ruling out event sequences that violate the communication model.  
The definitions below are those already provided in Section \ref{sec:backRel}. For example, 
for $CM1$ the constraint is that immediately after each sending event in $u$ we have its corresponding receiving event, with nothing in the middle; etc.

\begin{eqnarray*}
\lang{CM1}{\events} &=& 
  \{\eventtrace \in \langevents | \forall_{a\transmsg{M_1}b\in\interactions}.\forall_{k\in dom(\eventtrace)}. \send{a}{M_1} = \eventtrace[k-1] \implies 
  \\ & & \recv{b}{M_1} = \eventtrace[k] \} \\
  \lang{CM2}{\events} &=& 
 \{\eventtrace \in \langevents | \forall_{a\transmsg{M_1}b\in\interactions}.\forall_{c\transmsg{M_2}d\in\interactions}.\forall_{i,j,k,l\in dom(\eventtrace)}.
 \\ & & \recv{b}{M_1} = \eventtrace[i] \wedge \recv{d}{M_2} = \eventtrace[j] \wedge \send{a}{M_1} = \eventtrace[k] \wedge 
 \\ & & \send{c}{M_2} = \eventtrace[l] \wedge k < l \implies i < j\} \\
 \lang{CM3}{\events} &=& 
 \{\eventtrace \in \langevents | \forall_{a\transmsg{M_1}b\in\interactions}.\forall_{a\transmsg{M_2}d\in\interactions}.\forall_{i,j,k,l\in dom(\eventtrace)}.
 \\ & & \recv{b}{M_1} = \eventtrace[i] \wedge \recv{d}{M_2} = \eventtrace[j] \wedge \\ & &  \send{a}{M_1} = \eventtrace[k] \wedge \send{a}{M_2} = \eventtrace[l] \wedge k < l \implies i < j\}
 \end{eqnarray*}
\begin{eqnarray*}
\lang{CM4}{\events} &=& 
\{\eventtrace \in \langevents | \forall_{a\transmsg{M_1}b\in\interactions}.\forall_{c\transmsg{M_2}b\in\interactions}.\forall_{i,j,k,l\in dom(\eventtrace)}.
\\ & & \recv{b}{M_1} = \eventtrace[i] \wedge \recv{b}{M_2} = \eventtrace[j] \wedge \send{a}{M_1} = \eventtrace[k] \wedge \\ & &  \send{c}{M_2} = \eventtrace[l] \wedge k < l \implies i < j\}
\\
\lang{CM5}{\events} &=& 
\{\eventtrace \in \langevents | \forall_{a\transmsg{M_1}b\in\interactions}.\forall_{a\transmsg{M_2}b\in\interactions}.\forall_{i,j,k,l\in dom(\eventtrace)}.
\\ & & \recv{b}{M_1} = \eventtrace[i] \wedge \recv{b}{M_2} = \eventtrace[j] \wedge \send{a}{M_1} \causalrel{\eventtrace} \send{a}{M_2} \\ & &  \implies i < j\}
\\
& & \mbox{where } \send{a}{M_1}\causalrel{u}\send{b}{M_2} \iff\\
& & \hspace*{1cm}( (a = b \lor M_1=M_2) \wedge \\
& & \hspace*{1.2cm} \exists_{i,j \in dom(u)}.(u[i]=\send{a}{M_1} \wedge \send{b}{M_2}=u[j] \wedge i < j))  \\
& &  \hspace*{1cm} \vee \hspace*{0.2cm}(\exists_{ev \in \eventtrace}.\send{a}{M_1}\causalrel{u}ev\wedge ev\causalrel{u}\send{b}{M_2}) 
\\
\lang{CM6}{\events} &=&  \langevents 
\end{eqnarray*}

We can then apply a particular communication model to an \emph{interaction} protocol $\tau_i$ using  $\semAsync{\tau_i}{\moi}{\commodel}$, 
and to a \emph{message} protocol $\tau_m$ using $\semAsync{\tau_m}{}{\commodel}$,
which are defined as follows:
\begin{eqnarray*}
\semAsync{\tau_i}{\moi}{\commodel} &=& \semAsync{\tau_i}{\moi}{}\cap\lang{\commodel}{\eventsin{\interactionsin{\tau}}} \\
\semAsync{\tau_m}{}{\commodel} &=& \semAsync{\tau_m}{}{}\cap\lang{\commodel}{\eventsin{}(\tau)}
\end{eqnarray*}

\paragraph{Projection.}\label{sec:projection}

\newcommand{\distrib}[1]{\ulcorner{#1}\urcorner}

Projection is defined, intuitively, as focussing on the aspects of the protocol that are relevant for a given role. It is defined  as follows, where we write $\tau^A$ to denote projecting trace  $\tau$ for role $A$. 
\begin{eqnarray*}
(\emptyseq)^A &=& \emptyseq \\
(\atomicInt{a}{M}{b})^A &=& \send{a}{M}, \mbox{if } a=A \\ 
 &=& \recv{b}{M}, \mbox{if } b=A\\
 &=& \emptyseq, \mbox{otherwise} \\
(\send{a}{M})^A &=& \mbox{if } a=A \mbox{ then } \send{a}{M} \mbox{ else } \emptyseq \\
(\recv{a}{M})^A &=& \mbox{if } a=A \mbox{ then } \recv{a}{M} \mbox{ else } \emptyseq \\
(\tau_1 \otimes \tau_2)^A &=& (\tau_1)^A \otimes (\tau_2)^A \\
& & \mbox{Where $\otimes$ is any operator.}
\end{eqnarray*}
We then define the \emph{distribution} of $\tau$, denoted $\distrib{\tau}$, where $\tau$ involves roles $a_1 \ldots a_n$ as\footnote{We use $\|$ to distinguish between parallel composition of different agents, and parallel composition within a protocol. This distinction is used later in this section.}: 
\begin{eqnarray*}
\distrib{\tau} &=& \tau^{a_1} \| \ldots \| \tau^{a_n}
\end{eqnarray*}

To make an example, let us consider again the scenario proposed in Section \ref{sec:intro}. Alice decided to book Carol's apartment and now Carol needs some pieces of information from Alice in order to complete the reservation. This information can be wrong or incomplete, and Carol might need to ask Alice twice or more times. This can be represented using a cyclic specification
$$\mathit{reqInfo} = Alice \transmsg{Info} Carol ~\cdot~  
$$
$$(Carol \transmsg{Wrong} Alice ~\cdot~ \mathit{reqInfo} ~\lor~ Carol\transmsg{Booked}Alice)$$
where if the information provided by Alice is not satisfactory, Carol tells Alice and asks for new one (recursion on $\mathit{reqInfo}$). Once Carol will be satisfied with  Alice' answer, she will confirm the booking.
Thanks to cyclic specifications, we can represent protocols with infinite behaviours.
Let us consider $\mathit{main}$ as the combination of the two protocols: 
$\mathit{main} = \mathit{modifyRes} ~\cdot~ \mathit{reqInfo}$.

The projection of $\mathit{main}$ on each single agent would generate
\begin{eqnarray*}
\distrib{\mathit{main}} &=& \mathit{main}^{Alice} \;\|\; \mathit{main}^{Bob} \;\|\; \mathit{main}^{Carol}
\end{eqnarray*}
\begin{eqnarray*}
\mathit{main}^{Alice} &=& \mathit{modifyRes}^{Alice} ~\cdot~ \mathit{reqInfo}^{Alice}\\
\mathit{modifyRes}^{Alice} &=& \send{Alice}{Canc} ~\cdot~ \send{Alice}{Res}\\
\mathit{reqInfo}^{Alice} &=& \send{Alice}{Info} ~\cdot~ \\
& & (\recv{Alice}{Wrong} ~\cdot~ \mathit{reqInfo}^{Alice} ~\lor~ \recv{Alice}{Booked})
\end{eqnarray*}
\begin{eqnarray*}
\mathit{main}^{Bob} &=& \mathit{modifyRes}^{Bob} ~\cdot~ \mathit{reqInfo}^{Bob}\\
\mathit{modifyRes}^{Bob} &=& \recv{Bob}{Canc}\\
\mathit{reqInfo}^{Bob} &=& \epsilon
\end{eqnarray*}
\begin{eqnarray*}
\mathit{main}^{Carol} &=& \mathit{modifyRes}^{Carol} ~\cdot~ \mathit{reqInfo}^{Carol}\\
\mathit{modifyRes}^{Carol} &=& \recv{Carol}{Res}\\
\mathit{reqInfo}^{Carol} &=& \recv{Carol}{Info} ~\cdot~ \\
& & (\send{Carol}{Wrong} ~\cdot~ \mathit{reqInfo}^{Carol} ~\lor~ \send{Carol}{Booked})
\end{eqnarray*}

In order to define the semantics of a projected protocol we need to first define 
 what we term a \emph{decision structure}. This is needed in the semantics in order to deal correctly with projected protocols.  
Specifically, the intuition for enactability (see Section~\ref{sec:enactability})  is that an interaction protocol $\tau$ involving, say, three roles $a$, $b$ and $c$ is enactable iff there exist three protocols $\tau^a$, $\tau^b$ and $\tau^c$ such that their concurrent interleaving results in the same behaviour as the original protocol.
However, when a protocol contains choices ($\lor$) we need to ensure that the occurrences of $\lor$ in each of $\tau^a$, $\tau^b$ and $\tau^c$ arising from the same $\lor$ in $\tau$ are treated consistently.  For example, consider the protocol $\tau = a\transmsg{M_1}b \lor a\transmsg{M_2}c$. This protocol is simple: it specifies that agent $a$ can either send a message (``$M_1$'') to $b$, or it can send a different message (``$M_2$'') to agent $c$. When we distribute the protocol by projecting it (see Section~\ref{sec:projection}) and forming $\tau^a \| \tau^b \| \tau^c$ we obtain the distributed protocol
$ (\send{a}{M_1} \lor \send{a}{M_2}) \| (\recv{b}{M_1} \lor \varepsilon) \| (\varepsilon \lor \recv{c}{M_2}) $.
However, if we interpret each $\lor$ independently (as the semantics would naturally do) then we can have \emph{inconsistent} choices. For example, we could have $ (\send{a}{M_1}) \| (\varepsilon) \| (\varepsilon) $ where the message is sent by $a$, but $b$ does not elect to receive it. 
So what we need to do is ensure that each of the three occurrences of ``$\lor$'' represent the \emph{same} choice, and that the choice should be made consistently. 

The heart of the issue is that the trace expression notation offers a choice operator ($\lor$), which is adequate for global protocols. However, for local protocols it is important to be able to distinguish between a choice that represents a free (local) choice, and a choice that is forced by earlier choices. In this example, $a$ can freely choose whether to send $M_1$ or $M_2$. However, the choice of $b$ whether to receive $M_1$ or not is not a free choice, but is forced by $a$'s earlier choice.

Our semantics handles this by defining a \emph{decision structure} which is used to enforce consistent choices. 
Formally, given a protocol $\tau$ we define $d(\tau)$ as a set of \emph{decision structures} (formal definition below). A decision structure is a syntactic structure that mirrors the structure of $\tau$, except that each $\lor$ is annotated with a decision (e.g.~$L$ or $R$).
We define three operations defined on a decision structure: to get the sub-decision structure corresponding to the left part (denoted $d.L$), to get the right part ($d.R$) and to get the decision (L or R) associated with the current $\lor$ node (denoted $d.D$). We define $d(\tau)$ to create a set of decision structures, each of which corresponds to the structure of $\tau$, but where all possible assignments of decisions are made. 
Observe that If $\tau$ contains $N$ occurrences of $\lor$ then the set $d(\tau)$ contains $2^N$ elements. 
For example, given $\tau = \atomicInt{a}{M_1}{b} \lor \atomicInt{a}{M_2}{b}$   we have that $ d(\tau) = \{ \_ \overset{L}{\lor} \_, \_ \overset{R}{\lor} \_ \} $ where we use $\_$ to indicate an irrelevant part of a decision structure, and $\overset{L}{\lor}$ to denote a node tagged with a decision $L$. 

In addition to decisions of $L$ and $R$, the definition of $d(\tau_1 \lor \tau_2)$ has a second case ($\ldots \cup \{ t_1 \overset{LR}{\lor} t_2 \ldots$). The reason is that it is only possible to enforce consistent choice if the choice is made by a single agent. If this is not the case, then we annotate with ``$LR$'' to indicate that a mixed choice is possible.
For example, given $\tau = \atomicInt{b}{M_1}{a} \lor \atomicInt{a}{M_2}{b}$  we have that $ d(\tau) = \{ \_ \overset{LR}{\lor} \_ \} $ because $\ags{\tau_1} = \{b\} \neq \ags{\tau_2} = \{a\}$.
\begin{eqnarray*}
d(\varepsilon) &=& \{\varepsilon\} \\
d(I) &=& \{I\} \\
d(\tau_1 \lor \tau_2) &=& \{t_1 \overset{x}{\lor} t_2 \,|\, t_1 \in d(\tau_1) \land t_2 \in d(\tau_2)  \\ & & \hspace*{3mm} {} \land x \in \{R,L\} \land \ags{\tau_1} = \ags{\tau_2} \land |\ags{\tau_1}|=1 \} \\
 & & {} \cup \{t_1 \overset{LR}{\lor} t_2 \,|\, t_1 \in d(\tau_1) \land t_2 \in d(\tau_2)  \\ & &  \hspace*{3mm} {} \land ( (\ags{\tau_1} \neq \ags{\tau_2}) \lor (|\ags{\tau_1}| \neq 1)) \} \\
 & & \mbox{where } \ags{\tau} = \{p \; | \; p \transmsg{M} r \in \min(\mathrm{poset}(\tau)) \} \\
d(\tau_1 \oplus \tau_2) &=& \{t_1 \oplus t_2 \; | \;  t_1 \in d(\tau_1) \land t_2 \in d(\tau_2)\} \\ 
(\tau_L \otimes \tau_R).L &=& \tau_L \hspace*{0.85cm}
(\tau_L \otimes \tau_R).R \; \; = \; \; \tau_R \\
(\tau_L \overset{X}{\lor} \tau_R).D &=& X
\end{eqnarray*}
Where $\otimes$ is any operator, and $\oplus$ is any operator other than $\lor$.

We now specify the semantics of a distributed protocol, denoted $\sem{\tau}_{\mathrm{dist}}$. 
The semantics is defined in terms of a union over possible decision structures (first line).
The remaining of the equations for the semantics carry along the decision structure, and follow it in recursive calls, and for the semantics of $\lor$ it enacts the decision specified in the structure, rather than considering both sub-protocols.
Note that projection is defined using $\|$ rather than the usual $|$ - this differs in the semantics below, in that $\|$ passes the \emph{same} decision structure to both arguments. This ensures consistency between agents, but not within agents.
\begin{eqnarray*}
\sem{\tau}_{\mathrm{dist}} &=& \bigcup_{dt \in d(\tau)} \sem{\tau^{a_1} \| \ldots \| \tau^{a_n}}^{dt} \\
\sem{M}^{dt} &=& \{\langle M \rangle \} \\
\sem{\varepsilon}^{dt} &=& \{ \langle \rangle \} \\
\sem{\tau_1 \cdot \tau_2}^{dt} &=& \{ t_1 \circ t_2 | t_1 \in \sem{\tau_1}^{dt.L} \land t_2 \in \sem{\tau_2}^{dt.R} \} \\
\sem{\tau_1 \land \tau_2}^{dt} &=& \sem{\tau_1}^{dt.L} \cap \sem{\tau_2}^{dt.R} \\
\sem{\tau_1 \lor \tau_2}^{dt} &=& \mbox{if } dt.D=R \mbox{ then } \sem{\tau_2}^{dt.R} \\
& & \mbox{ elseif } dt.D=L \mbox{ then } \sem{\tau_1}^{dt.L}\\
& &  \mbox{ else } \sem{\tau_2}^{dt.R}  \cup \sem{\tau_1}^{dt.L} \\
\sem{\tau_1 | \tau_2}^{dt} &=& \{ z |  t_1 \in \sem{\tau_1}^{dt.L} \land t_2 \in \sem{\tau_2}^{dt.R} \land z \in t_1 \bowtie t_2\} \\
\sem{\tau_1 \| \tau_2}^{dt} &=& \{ z |  t_1 \in \sem{\tau_1}^{dt} \land t_2 \in \sem{\tau_2}^{dt} \land z \in t_1 \bowtie t_2\} 
\end{eqnarray*}
Where $t_1 \bowtie t_2$ is the set of all interleavings of  $t_1$ and $t_2$, and $\circ$ is concatenation over sequences.
Note that if $\tau$ does not contain any occurrences of  $\lor$ then the semantics above reduce to the standard semantics.  

Finally, we define 
 $\semAsync{\tau_i}{\mathrm{dist}}{\commodel}$, which computes the semantics of an interaction protocol $\tau_i$ by distributing it, and also applies a particular communication model $\commodel$. 
\begin{eqnarray*}
\semAsync{\tau_i}{\mathrm{dist}}{\commodel} &=& \semAsync{\tau_i}{\mathrm{dist}}{}\cap\lang{\commodel}{\eventsin{\interactionsin{\tau}}} 
\end{eqnarray*}

\paragraph{Enactability.}\label{sec:enactability}

\begin{figure*}[!t]
\begin{scriptsize}
\begin{tabular}{|l|l|l|l|l|}
\hline \multicolumn{5}{|c|}{$\atomicInt{a}{M_1}{b} ~\cdot~ \atomicInt{b}{M_5}{c}$} \\\hline CM&RS&RR&SS&SR\\\hline
CM1 & \yes & \yes & \yes & \yes\\
CM2 & \yes & (\yes) & (\yes) & (\yes)\\
CM3 & \yes & (\yes) & (\yes) & (\yes)\\
CM4 & \yes & (\yes) & (\yes) & (\yes)\\
CM5 & \yes & (\yes) & (\yes) & (\yes)\\
CM6 & \yes & (\yes) & (\yes) & (\yes)\\
\hline
\end{tabular}
\enskip
\begin{tabular}{|l|l|l|l|l|}
\hline \multicolumn{5}{|c|}{$\atomicInt{a}{M_1}{b} ~\cdot~ \atomicInt{a}{M_2}{c}$} \\\hline CM&RS&RR&SS&SR\\\hline
CM1 & \yes & \yes & \yes & \yes\\
CM2 & \no & \yes & \yes & (\yes)\\
CM3 & \no & \yes & \yes & (\yes)\\
CM4 & \no & \no & \yes & (\yes)\\
CM5 & \no & \no & \yes & (\yes)\\
CM6 & \no & \no & \yes & (\yes)\\
\hline
\end{tabular}
\enskip
\begin{tabular}{|l|l|l|l|l|}
\hline \multicolumn{5}{|c|}{$\atomicInt{a}{M_1}{b} ~\cdot~ \atomicInt{c}{M_6}{b}$} \\\hline CM&RS&RR&SS&SR\\\hline
CM1 & \yes & \yes & \yes & \yes\\
CM2 & \no & \yes & \yes & (\yes)\\
CM3 & \no & \yes & \no & (\yes)\\
CM4 & \no & \yes & \yes & (\yes)\\
CM5 & \no & \yes & \no & (\yes)\\
CM6 & \no & \yes & \no & (\yes)\\
\hline
\end{tabular}
\enskip
\begin{tabular}{|l|l|l|l|l|}
\hline \multicolumn{5}{|c|}{$\atomicInt{a}{M_1}{b} ~\cdot~ \atomicInt{c}{M_4}{a}$} \\\hline CM&RS&RR&SS&SR\\\hline
CM1 & \yes & \yes & \yes & \yes\\
CM2 & \no & \no & \no & \yes\\
CM3 & \no & \no & \no & \yes\\
CM4 & \no & \no & \no & \yes\\
CM5 & \no & \no & \no & \yes\\
CM6 & \no & \no & \no & \yes\\
\hline
\end{tabular} 
\\[3mm]
%
\begin{tabular}{|l|l|l|l|l|}
\hline \multicolumn{5}{|c|}{$\atomicInt{a}{M_1}{b} ~\cdot~ \atomicInt{a}{M_2}{b}$} \\\hline CM&RS&RR&SS&SR\\\hline
CM1 & \yes & \yes & \yes & \yes\\
CM2 & \no & \yes & \yes & (\yes)\\
CM3 & \no & \yes & \yes & (\yes)\\
CM4 & \no & \yes & \yes & (\yes)\\
CM5 & \no & \yes & \yes & (\yes)\\
CM6 & \no & (\yes) & (\yes) & (\yes)\\
\hline
\end{tabular}
\enskip
\begin{tabular}{|l|l|l|l|l|}
\hline \multicolumn{5}{|c|}{$\atomicInt{a}{M_1}{b} ~\cdot~ \atomicInt{b}{M_3}{a}$} \\\hline CM&RS&RR&SS&SR\\\hline
CM1 & \yes & \yes & \yes & \yes\\
CM2 & \yes & (\yes) & (\yes) & (\yes)\\
CM3 & \yes & (\yes) & (\yes) & (\yes)\\
CM4 & \yes & (\yes) & (\yes) & (\yes)\\
CM5 & \yes & (\yes) & (\yes) & (\yes)\\
CM6 & \yes & (\yes) & (\yes) & (\yes)\\
\hline
\end{tabular}
\enskip
\begin{tabular}{|l|l|l|l|l|}
\hline \multicolumn{5}{|c|}{$\atomicInt{a}{M_1}{b} ~\lor~ \atomicInt{a}{M_2}{c}$} \\\hline CM&RS&RR&SS&SR\\\hline
CM1 & \yes & \yes & \yes & \yes\\
CM2 & \yes & \yes & \yes & \yes\\
CM3 & \yes & \yes & \yes & \yes\\
CM4 & \yes & \yes & \yes & \yes\\
CM5 & \yes & \yes & \yes & \yes\\
CM6 & \yes & \yes & \yes & \yes\\
\hline
\end{tabular}
\enskip
\begin{tabular}{|l|l|l|l|l|}
\hline \multicolumn{5}{|c|}{$\atomicInt{a}{M_1}{b} ~\lor~ \atomicInt{b}{M_3}{a}$} \\\hline CM&RS&RR&SS&SR\\\hline
CM1 & \yes & \yes & \yes & \yes\\
CM2 & \no & \no & \no & \no\\
CM3 & \no & \no & \no & \no\\
CM4 & \no & \no & \no & \no\\
CM5 & \no & \no & \no & \no\\
CM6 & \no & \no & \no & \no\\
\hline
\end{tabular}
\caption{Automatically generated analyses of enactability}\label{figtable2}\label{figtable1}
\end{scriptsize}
\end{figure*}

We are now finally in a position to define enactability. The intuition is that an interaction protocol $\tau$ is enactable iff the semantics of $\tau$, with respect to a selected message ordering interpretation and communication model, can be realised by a distributed version of the protocol. In other words, if there exists for each role $r$ a corresponding message protocol $\tau_r$ such that the combination of these protocols realises the same behaviour as $\tau$. 
However, instead of considering whether there exists some $\tau_r$, we let $\tau_r = \tau^r$, i.e.~we take for each role the projected protocol as its protocol. 

We also consider a notion of \emph{weak} enactability. This applies in a situation where the a distributed enactment is able to avoid violating the behaviour specified by $\tau$, but is not able to recreate all of the behaviours that $\tau$ specifies. 
This situation can arise with weaker message ordering interpretations (see below for examples). 
Weak enactability can also arise in situations where two ordered messages have two overlapping roles (e.g. $\tau = \atomicInt{a}{M_1}{b} \cdot \atomicInt{b}{M_2}{a}$). In this situation the projection operator is too strict: it has $\tau^b = r(M_1) \cdot s(M_2)$, but if we adopt an SR message ordering interpretation, then we do not need to ensure that $M_2$ is sent after $M_1$ is received, only that $M_1$ is sent before $M_2$ is received, which role $a$ can ensure on its own. 

\begin{definition}[Strongly/Weakly Enactable]
\label{sw-enact-async-def}
Let $\tau$ be an interaction protocol, $\{ a_1, a_2,$ $...,$ $a_n \}$ the set of agents involved in $\tau$, $\moi\in\moiset$ a message order interpretation and $\commodel\in\commodelset$ a communication model. We say that, $\tau$ is strongly (weakly) enactable, for $\moi$ semantics in $\commodel$ model iff the decomposition of $\tau$ through projection on its agents $\{ a_1,a_2,...,a_n \}$ recognizes the same (a subset of) traces recognized by $\tau$.
Formally: 
\begin{eqnarray*}
\mathit{enact}(\tau)^{\commodel}_{\moi}  & \mbox{iff} & 
	 \semAsync{\tau}{\mathrm{dist}}{\commodel} = \semAsync{\tau}{\moi}{\commodel} \\
\mathit{weak\_enact}(\tau)^{\commodel}_{\moi} & \mbox{iff} & 
	 \semAsync{\tau}{\mathrm{dist}}{\commodel} \subseteq \semAsync{\tau}{\moi}{\commodel} 
\end{eqnarray*}
\end{definition}

If a protocol is weak enactable, the interleaving of the corresponding local protocols generates a subset of its traces (with a fixed moi and communication model). In practice, this means that our implementation is sound (generates only valid traces), but it is not complete (not all the traces are generated). Consequently, our system will be more restrictive than we wanted.

Figure \ref{figtable2} show the results of applying this definition to a number of cases, with different message ordering interpretation, and different communication models. These tables were all generated by the Haskell implementation of the definitions in this paper, in which \yes and (\yes) denote \emph{strongly} and \emph{weakly} enactable, respectively.
The prototype counts \textasciitilde300 LOC. It implements the trace expression standard semantics, message order interpretation, communication model semantics and enactability check\footnote{The code is available on the web at:
\url{http://enactability.altervista.org/}}. 

Looking at the tables in Figure~\ref{figtable1}, we make the following observations. 

Firstly, CM1 is quite strict: all the cases considered are enactable under CM1, regardless of the selected message ordering interpretation. This is expected: we know that CM1 is quite strong.

Secondly, for many examples there is not a difference in enactability with the different communication models (other than CM1), except where the communication model corresponds to the combination of MOI and the pattern in the protocol. For example, in the top row, second table from the right, the simple protocol is enactable given SS message ordering interpretation only with CM2 and CM4 (and, of course, CM1). This is because for this protocol both messages are received by the same agent but sent by different agents, and, given an RR MOI, the desired constraint that agent $B$ receives the first message before the second, can only be enforced using a communication model that guarantees delivery of messages to the same recipient in the order in which messages were sent. Both CM2 and CM4 provide this guarantee (in fact CM4 provides exactly this, and CM2 is stronger).

Thirdly, RS appears to be a good choice for message ordering interpretation, since it is the only MOI where protocols are never weakly enactable. For the other message ordering interpretations, there are protocols that are only weakly enactable (for communication models other than CM1). A protocol being weakly enactable indicates that the desired behaviour specified by the MOI is too loose: it permits behaviours that the distributed realisation cannot realise. 
On the other hand, in the case of the left-most table on the bottom row (protocol $\atomicInt{a}{M_1}{b} ~\cdot~ \atomicInt{a}{M_2}{b}$), the protocol is not enactable under RS (except for CM1), but is enactable under SS and under RR.
Turning to SR, we observe that it seems to be too weak: almost all the protocols in the figure are enactable (although in most cases only weakly enactable). 

Returning to the example from the introduction:
$$\mathit{modifyRes} = Alice \transmsg{Canc} Bob ~\cdot~  Alice \transmsg{Res} Carol$$ where $a1 \transmsg{M} a2$
this example corresponds to the second table from the left in the top row of  Figure~\ref{figtable1}. This shows that, if one desires an $RR$ MOI, i.e.~that what is meant by $Canc$ coming before $Res$ is that Bob receives the $Canc$ message before Carol receives the $Res$ message, then the underlying message communication must be $CM1$, $CM2$ or $CM3$, in order for the protocol to be enactable.

\section{Discussion}\label{sec:relatedwork}

Despite the large amount of work on enactability, very few approaches consider how message ordering and decision structures affect its definition, very few come with an implemented prototype, and none considers the issues raised by the communication model. 

Although one motivation might be that it is generally desirable to have robust protocol specifications that are independent of the underlying platform implementation, also ensuring separation of concerns, we observe that robustness could make the protocol too complex, or harder to maintain. Considering what the underlying implementation guarantees w.r.t. communication model, we can relax our specifications, and above all, a protocol that is not enactable in some platform, can be in some other. This makes our work relevant to platform designers, and protocol designers. 

Taking all these features into account in a unified semantic-driven way, and demonstrating the potential of the approach on a highly expressive protocol language, are the innovative and original features of this contribution. 

Desai and Singh \cite{DBLP:conf/aaai/DesaiS08} limit their investigation to the RS message ordering interpretation, that they consider the standard of correctness. Hence, despite the nice introduction they provide to other message orderings and to the problems they might raise, the definition of enactability they provide is not parametric in the MOI. 

Lanese et al. \cite{DBLP:conf/sefm/LaneseGMZ08} move a step further, but the generality of their approach is still limited. They define three different notions of enactability, that they name conformance: sender conformance, receiver conformance, and disjoint conformance. That approach is more flexible that the one by Desai and Singh, but less general than ours, where the definition of enactability is parametric in the MOI and does not require different cases. Also, they only consider how sequence and choice are affected by MOIs, leaving the study of other operators for the future. Moreover, when discussing interaction protocols whose most external operator is a choice, they put a very strong constraint for enactability, namely that the agents involved in the two branches of the choice (excluding the agents involved in the choice itself) are the same. We added decision structures to overcome this restriction, and provide a notion of enactability that can succeed even when that constraint is not met.

Neither Desai and Singh, nor Lanese et al., use formalisms for protocol representation as expressive as trace expressions, and neither of them presents experiments obtained from a working prototype, as we do. 

With respect to the introduction of decision structures to remove unnecessary restrictions on enactability of protocols when choice is involved, our proposal is similar to that by Qiu et al., \cite{DBLP:conf/www/QiuZCY07}, as for the other works we have discussed in this section, we implemented our enactability checker, whereas their work  only provides definitions. Additionally, our approach is simpler in that we do not need to label the choice operator with agents as they do. 

In the future, we will address both theoretical and practical issues. 
On the theoretical side, we will carry out a systematic analysis of the relationships between Communication Model and Message Ordering Interpretation, to identify those combinations which provide some guarantees by design.  
We will also consider the relationships between enactability and distributed monitorability \cite{DBLP:conf/atal/FerrandoAM17}, as they might turn out to resort to the same definition.

On the practical part, we plan to improve our working prototype to provide a useful tool to assess protocols for enactability. Apart from providing a user-friendly interface, a key issue to address will be to provide a way to isolate the part of a non-enactable protocol that makes it non-enactable.
Also, trace expressions are interpreted in a coinductive way \cite{Sangiorgi:2009:OBC:1516507.1516510} to represent infinite traces of events.
Since Haskell does not support coinduction, the existing prototype can be only used on acyclic message and interactions protocols. Haskell has been chosen because the implementation mimics the semantics requiring next to no effort. In order to fully implement the proposed features we are planning to develop the enactability check using SWI-Prolog\footnote{\url{http://www.swi-prolog.org}}, which natively supports coinduction.
To stress-test the prototype and assess its performance from a qualitative and quantitative viewpoint we plan to create a library of  interaction protocols known to be ``problematic'' w.r.t. enactability, and perform systematic experiments. 

Finally, this work highlighted the need of characterising the existing agent infrastructures like Jade \cite{JadeBook}, Jason \cite{Bordini:2007:PMS:1197104}, Jadex \cite{Pokahr2005}, etc, in terms of the communication model they support. This would allow us to state if a protocol is enactable on a given infrastructure, strengthening the potential of our proposal to be exploited in real applications.

\bibliographystyle{unsrt}  
\bibliography{references}  

\end{document}